\documentclass[
    reprint, 
    aps,
    preprintnumbers,
    twocolumn,
    prb,
    superscriptaddress
]{revtex4-2}

\usepackage[utf8]{inputenc}
\usepackage{booktabs}
\usepackage[multiple]{footmisc}
\usepackage{lipsum}
\usepackage{rotating}
\usepackage{perpage}
\usepackage{chronology}
\usepackage{lipsum}
\usepackage{amssymb}
\usepackage{amsbsy}
\usepackage{amsmath}
\usepackage{tikz}
\usepackage[T1]{fontenc}
\usepackage{etoolbox}
\usepackage{graphics}
\usepackage{abraces}
\usepackage{siunitx}
\usepackage{hyperref}
\usepackage{multirow}
\usepackage{mathtools}
\usepackage{bm}
\usepackage{url}
\usepackage{physics}
\usepackage{hyperref}
\usepackage{bbm}
\usepackage{color}
\usepackage{placeins}
\usepackage[normalem]{ulem}

\def\dul#1{\underline{\underline{#1}}}
\def\enforce#1{$\text{#1}$}

\newcommand{\mb}{\mathbf}
\newcommand{\operator}[1]{\mathcal{#1}}

\begin{document} 

\title{Understanding the dynamics of randomly positioned dipolar spin ensembles}


\author{Timo Gräßer}
\email{timo.graesser@tu-dortmund.de}
\affiliation{Condensed Matter Theory, TU Dortmund University,
Otto-Hahn Stra\ss{}e 4, 44221 Dortmund, Germany}

\author{Kristine Rezai}
\email{kristinerezai@gmail.com}
\affiliation{Department of Physics, Harvard University, Cambridge, MA 02138, USA}
\affiliation{Department of Physics, Boston University, Boston, MA 02215, USA}

\author{Alexander O.~Sushkov}
\email{asu@bu.edu}
\affiliation{Department of Physics, Boston University, Boston, MA 02215, USA}

\author{G\"otz S.~Uhrig}
\email{goetz.uhrig@tu-dortmund.de}
\affiliation{Condensed Matter Theory, TU Dortmund University,
Otto-Hahn Stra\ss{}e 4, 44221 Dortmund, Germany}

\date{\today}

    \begin{abstract}
    Dipolar spin ensembles with random spin positions attract much attention currently
		because they help to understand decoherence as it occurs in solid state quantum bits
		in contact with spin baths. Also, these ensembles are systems which 
		may show many-body localization, at least
		in the sense of very slow spin dynamics. 
		We present measurements of the autocorrelations
		of spins on diamond surfaces at infinite temperature in a doubly-rotating frame which eliminates local disorder.
		Strikingly, the time scales in the longitudinal and the transversal channel differ
		by more than one order of magnitude which is a factor much greater than one would have expected
		from simulations of spins on lattices. A previously developed dynamic mean-field theory 
		for spins (spinDMFT) fails to explain this phenomenon. Thus, we improve it by extending 
		it to clusters (CspinDMFT). This theory does capture the striking mismatch up to two orders of 
		magnitude for random ensembles.	Without positional disorder, however, the mismatch is only moderate with 
		a factor below 4. The pivotal role of positional disorder suggests that the strong mismatch
		is linked to precursors of many-body localization.
		\end{abstract}

\maketitle

\section{Introduction}
\label{s:intro}

Understanding and controlling the dynamics of many-body quantum systems enables the
development of novel materials and technologies for quantum devices for sensing, computation, communication, and modeling~\cite{niels00,ladd10b}.
Such systems have been realized in a variety of platforms: 
ultracold atoms~\cite{bloch08}, trapped ions~\cite{blatt12}, superconducting devices~\cite{koch07,baren14}, and 
solid state quantum bits (qubits)~\cite{doher13,schir14,sipah16,alvar15,wei18}. 
Access to the dynamics of a single qubit is especially important in heterogeneous 
systems~\cite{schli17,tamin14} where the local environment of the qubit plays a vital role and the averaging 
over an ensemble can conceal important features of the temporal evolution.
The interplay between the local environments, interactions, and the dimensionality of the system
has been the subject of extensive theoretical and experimental investigations~\cite{gopal16,burin06,yao14}.
The systems under study allowed for the experimental observations of nonequilibrium quantum dynamics in strongly interacting systems such as the propagation of quantum correlations~\cite{jurce14,riche14,parso16}, nonequilibrium phases of matter~\cite{zhang17c,choi17,keesl19,rispo19,ebadi21}, 
and the phenomena of relaxation and localization 
effects~\cite{alvar15,wei18,smith16,choi16,kucsk18,schre15,kaufm16,davis23}.

We consider the spin dynamics of two-dimensional ensembles of randomly positioned spins
with $S=1/2$ with long-range magnetic dipolar interactions at high temperature in the spin disordered state. Dipolar spin ensembles have attracted much attention in the last years due to their potentially very long coherence times \cite{nandk21,artia21,zu21,peng23,marti23}.
The considered spins have two different quantum states and 
thus can be viewed as quantum bits.
Such systems have been widely studied also in solid state NMR~\cite{carde14}
 and they were the starting point for Anderson's work on localization~\cite{ander58}. 
The spatial propagation of spin states takes place via flip-flop processes
at a rate set by the dipolar interaction strength $J$.
Varying local energies, for instance due to local intrinsic and extrinsic magnetic fields, suppress
this propagation due to a mismatch of energies at the involved sites.

To fully capture the dynamical interplay represents a formidable task.
Due to the exponentially growing dimension of the relevant Hilbert space with the number 
of spins, a direct, brute-force approach is out of the question. Exact diagonalization and Chebyshev polynomial expansion are restricted to 
fairly small sample sizes \cite{talez84,weiss06a}. Density-matrix renormalization works
for chains \cite{schol05} and star-topologies \cite{stane13} and is restricted in the maximum time
that can be reached. Semiclassical approaches related to the
truncated Wigner approximation \cite{chen07,polko10,stane14b,david15,zhu19} neglect by construction a large
amount of quantumness, in particular interference effects which matter
for the effects studied here. Approaches based on Master equations \cite{duboi21}
require that there is a priori a distinction of system and bath
including a clear separation in energy scales. Cluster expansions 
represent powerful tools \cite{witze05,witze06,lindo18,saiki07,yang08a,yang09a}, but their application gets cumbersome
in disordered systems. Finally, Monte Carlo approaches are not particularly 
efficient for real-time dynamics.

Thus, the first main objective of this article is to develop a tractable numerical 
approach for the local spin dynamics in a spin disordered state of the random spin ensemble.
To this end, we extend the dynamic mean-field approach for disordered spin systems
which we recently introduced under the name ``spinDMFT'' \cite{grass21}.
The extension takes the quantum dynamics of a local cluster of spins completely into
account and applies the concept of dynamical mean fields only to the surrounding
spins which are not part of the considered cluster; we suggest the
acronym CspinDMFT for the extended approach. Note the similarity of the basic
idea of this extension to the extension of fermionic DMFT \cite{georg96} to cellular
DMFT \cite{kotli01}. The need for a cluster extension of spinDMFT has been noted already
and carried out on the level of spin dimers \cite{marti23}.

Our second main objective is to understand the vastly differing time scales of
longitudinal and transversal relaxation,  $T_{1\rho}$ and $T_2$, respectively,
observed experimentally.
We use the $T_{1\rho}$ at strong drive as a measurement of longitudinal relaxation 
instead of the $T_1$ since slowdown of $T_1$ relaxation is already expected due to the 
presence of extrinsic disorder \cite{rezaiarxiv}.  However, the $T_{1\rho}$ at strong 
drive evolves only under dipolar spin dynamics as the effects of 
extrinsic disorder are reduced. We show that the observed
difference of one to two order of magnitude cannot be explained by spins
on regular lattices, but that irregular, random configurations are necessary
to account for the qualitative difference. 
This suggests that the
very long longitudinal times $T_{1\rho}$ may be the precursors of many-body localization \cite{nandk15,altma18,abani19}, at least
in the weak sense of the occurrence of very long time scales \cite{evers23}. 
We stress that the theoretical explanation for the strongly differing time scale
in the transversal and longitudinal channel required the methodological progress, i.e.,
passing from spinDMFT to CspinDMFT.

We study the experimental platform consisting of an ensemble of paramagnetic two-level systems (surface spins) 
on the diamond surface~\cite{grotz11,grino14}. These two-level systems stem from electronic 
spins $S=1/2$ and are likely localized defect states ~\cite{sangt19,stace19} on the surface of diamond.
A single, shallow NV center measures the dynamics of the system of surface spins.
While recent experimental studies have shown that some of the surface spins may be mobile depending on initial surface treatment~\cite{dwyer22}, we have verified that in our experiments, the central surface spin's position remains 
stable~\cite{rezaiarxiv}. 
Even if the surface spins moved and changed their positions between consecutive
measurements, the theoretical approach used in this article
would be justified even better once the simulated data are averaged.

\begin{figure}
    \centering
    \includegraphics{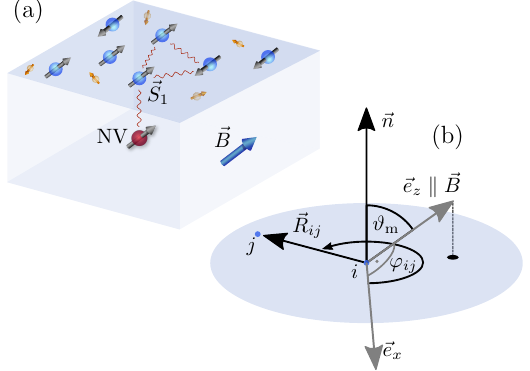}
    \caption{Experimental system and its geometry. (a) The experimental system consists of a single near-surface NV center in diamond strongly coupled to one surface electron spin $\vec{S_1}$, which interacts with other surface electron spins and a bath of proton nuclear spins. (b) Sketch of the geometry of the surface spin system. 
    Here, $\vec{n}$ is the normal vector of the surface and $\vec{B}$ the static magnetic field pointing into the $z$-direction. Since the spins are restricted to the surface, the distance vector $\vec{R}_{ij}$ can be expressed by surface polar coordinates $R_{ij}$ and $\varphi_{ij}$.
    }
    \label{fig:SurfacePolarCoordinates}
\end{figure}

Although the interaction of the surface spins is of dipolar nature, their dynamics under the experimentally applied fields is considerably different. For the \emph{singly-rotating frame} dynamics (with longitudinal relaxation time $T_1$ and and dephasing time $T_2$), we consider a strong static magnetic field $\vec{B}$ defining the 
$z$-direction. It encloses the magic angle $\vartheta_m$ with the normal vector of the diamond surface as indicated in Fig.~\ref{fig:SurfacePolarCoordinates}. Passing to the corresponding Larmor rotating frame and applying the rotating wave approximation leads to the effective interaction
\begin{align}
    \mb{H}_{\text{dd}}^{\text{s-rot}} &=
		\frac12 \sum_{i\neq j} J_{ij} \left( -\frac12 \mb{S}_i^x\mb{S}_j^x 
		- \frac12 \mb{S}_i^y\mb{S}_j^y + \mb{S}_i^z\mb{S}_j^z \right),
    \label{eqn:SRF_Hamiltonian}
\end{align}
which is anisotropic. The couplings are given by
\begin{align}
    J_{ij} &= - \frac{\hbar^2 \gamma^2}{R_{ij}^3} \cos 2 \varphi_{ij},
    \label{eqn:magic_angle_coupling}
\end{align} 
where $\varphi_{ij}$ is the angle between the distance vector 
$\vec{R}_{ij}~:=~\vec{R}_{i}-\vec{R}_{j}$ and the $x$-direction, see Fig.~\ref{fig:SurfacePolarCoordinates}.
For the \emph{doubly-rotating frame} dynamics ($T_{1\rho}$), we additionally consider a strong drive field perpendicular to $\vec{B}$ and rotating at the Larmor frequency. The direction of the drive defines the spin $y$-component. Starting from Eq.\ \eqref{eqn:SRF_Hamiltonian} and turning to 
another rotating frame including the application of  the rotating wave approximation results in
\begin{align}
    \mb{H}_{\text{dd}}^{\text{d-rot}} &=
		\frac12 \sum_{i\neq j} J_{ij} \left( \frac14 \mb{S}_i^z\mb{S}_j^z 
		+ \frac14 \mb{S}_i^x\mb{S}_j^x -\frac12 \mb{S}_i^y\mb{S}_j^y \right).
    \label{eqn:DRF_Hamiltonian}
\end{align}
Here, the distinguished direction is the $y$-direction which is henceforth called the longitudinal one. The other two directions are henceforth called the transversal ones. Note that there is a relative factor of 2 in the overall couplings compared to the singly-rotating frame Hamiltonian \eqref{eqn:SRF_Hamiltonian}. This needs to be considered, when comparing results from both reference systems with each other.

The article is structured as follows. In Sect.~\ref{sec:spinDMFT}, we apply spinDMFT to an 
inhomogeneous ensemble of dipolar spins in the doubly-rotating frame as in experiment. The striking
discrepancy leads us to Sect.~\ref{sec:ClusterspinDMFT}, in which we extend spinDMFT 
by treating clusters of spins fully  quantum-mechanically. The so-derived cluster spinDMFT
(CspinDMFT) is subsequently analyzed for its performance by investigating 
its convergence for isotropically coupled spins on a triangular lattice as well as on 
inhomogeneous graphs.  Section~\ref{sec:ApplicationClusterspinDMFT} is devoted to the 
application of CspinDMFT to the experimental scenario. We compare the decay times $T_2$ and 
$T_{1\rho}$ obtained from CspinDMFT to those measured in experiment.
Good agreement is found supporting the hypothesis that the positional disorder in the spin system
is at the origin of the strongly differing transversal and longitudinal decay times as
concluded in Sect.~\ref{sec:Conclusion}. In the appendices, we discuss experimental details, relevant numerical issues 
and analytical subtleties.

\section{Simulating dipolar spin dynamics with \enforce{spin}DMFT}
\label{sec:spinDMFT}

As a starting point we apply spinDMFT to the Hamiltonian \eqref{eqn:DRF_Hamiltonian} for
an inhomogeneous dipolar spin ensemble, i.e., for random positions of the spins. One realizes,
however, that the reduction to an effective single site problem oversimplifies the physical 
setting because it implies that only one energy scale, or equivalently time scale, matters
in the problem. The aspect of the random positions of the spins does not appear anymore.
The calculation for a perfectly ordered spin lattice with appropriate energy
scale would be identically the same.

The derivation of spinDMFT will be sketched for completeness, but 
we refer the reader to Ref.~\onlinecite{grass21} for its detailed justification.
The starting point is the Hamiltonian in Eq.~\eqref{eqn:DRF_Hamiltonian}. 
It can be rewritten in the form
\begin{align}
    \operator{H}_{\text{d-rot}} = \frac12 \sum_{i} \vec{\mb{S}}_i \cdot \vec{\mb{V}}_i
\end{align}
by introducing the operators for the local environments
\begin{align}
    \vec{\mb{V}}_i &\eqqcolon \sum_{j,j\neq i} J_{ij} \dul{D} \vec{\mb{S}}_{j}, & 
    \dul{D} &= 
    \begin{pmatrix}
        \frac14 & 0  & 0 \\
        0 & -\frac12 & 0 \\
        0 & 0  & \frac14 \\
    \end{pmatrix}.
    \label{eqn:DRF_localenvironments}
\end{align}
In spinDMFT, these operators are replaced by time-dependent random mean-fields drawn from normal distributions.
This leads to the time-dependent mean-field Hamiltonian
\begin{align}
    \mb{H}_{\text{d-rot},i}^{\text{mf}}(t) &= \vec{ \mb{S}}_{i} \cdot \vec{V}_{i}(t)
    \label{eqn:drotHamiltonian}
\end{align}
for the spin $i$ of the system along with the self-consistency conditions
\begin{subequations}
\begin{align}
    \overline{V_{i}^{x}(t)V_{i}^{x}(0)} &= \frac1{16} J_{\text{Q},i}^2\langle \mb{S}_{i}^{x}(t) \mb{S}_{i}^{x}(0) \rangle, \\
    \overline{V_{i}^{y}(t)V_{i}^{y}(0)} &= \frac1{4}  J_{\text{Q},i}^2\langle \mb{S}_{i}^{y}(t) \mb{S}_{i}^{y}(0) \rangle, \\
    \overline{V_{i}^{z}(t)V_{i}^{z}(0)} &= \frac1{16} J_{\text{Q},i}^2\langle \mb{S}_{i}^{z}(t) \mb{S}_{i}^{z}(0) \rangle,
\end{align}
\label{eqn:spinDMFTselfcons}
\end{subequations}
connecting the second mean-field moments to spin autocorrelations. This implies that 
the quadratic coupling constant
\begin{align}
    J_{\text{Q},i}^2 \eqqcolon \sum_{j} J_{ij}^2 = \sum_{j} \frac{\hbar^4 \gamma^4}{R_{ij}^6} \cos^2 2 \varphi_{ij},
\end{align}
captures the whole dependence of the spin-spin autocorrelations on the spatial distribution of spins. 
The anisotropy in the couplings yields the relative factor of $4$ between the 
longitudinal ($y$) and transversal ($x,z$) self-consistency equations. 

\begin{figure}[ht]
    \includegraphics[width=1.0\columnwidth]{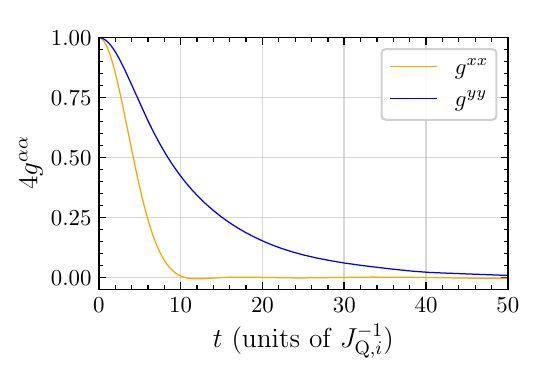}
    \caption{Universal spin autocorrelations in the doubly-rotating frame obtained in spinDMFT. The spatial distribution of spins enters only in the quantitative determination of the energy scale $J_{\text{Q},i}$ whose reciprocal value is the unit of time if $\hbar$ is set to one.}
    \label{fig:spinDMFT_DDRF}
\end{figure}

To solve the self-consistency problem by iteration we start from some fairly arbitrary initial guess 
for the local environment field of the considered spin to determine the spin-spin autocorrelation.
In the next iteration, the autocorrelations define the moments of the normal distribution for the local 
fields so that one obtains improved results for the autocorrelations. Note that one has
to average over a sufficiently large number of time-dependent local fields (about $10^5$ in our
calculations) to obtain
the relevant autorcorrelations with good accuracy. This step is repeated till
convergence is achieved within a tolerance of $\num{7e-4}$ for the
spin-spin autocorrelations
\begin{align}
    g^{\alpha\beta} (t) \eqqcolon \langle \mb{S}^{\alpha}(t) \mb{S}^{\beta}(0) \rangle,
\end{align}
which are plotted in Fig.~\ref{fig:spinDMFT_DDRF}. The result is universal in the sense
that only the energy (time) scale needs to be determined; no trace of the random
distribution of the spins in space enters anymore. There is roughly a factor
of 2 to 4 between the decay time in the longitudinal $yy$-channel and the one in the 
transversal $xx$-channel. This is in-line with the factor of 2
 in the anisotropic couplings in Eq.~\eqref{eqn:DRF_Hamiltonian}.

\begin{widetext}

\begin{figure}[bt]
    \includegraphics[width=\textwidth]{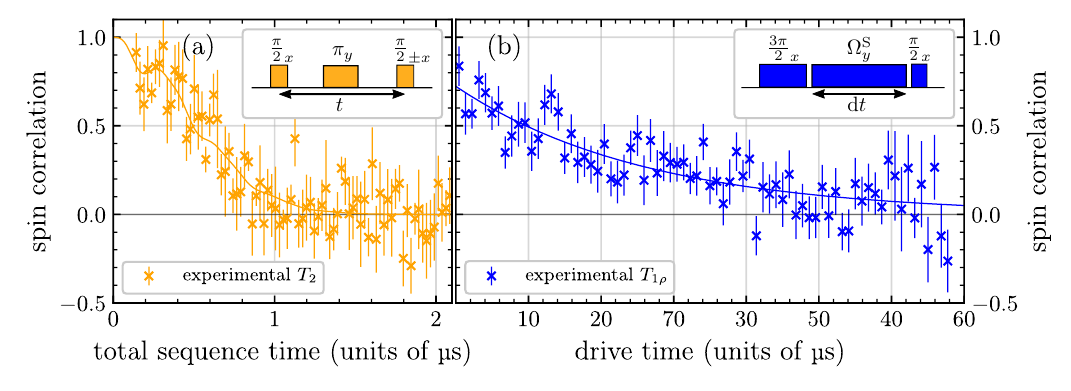}
    \caption{Measurement of the lifetime of the central spin state. 
		Panel (a): Surface spin Hahn echo 
		measurement of $\langle S_i^x(t) S_i^x{0}\rangle$, the transversal spin autocorrelation,
		used to determine $T_2$ along with the applied pulse sequence shown in the top right corner. Since the measurement takes place in the singly-rotating frame, the spin dynamics are affected by the proton nuclear spin bath at the surface. To display this, the data are fitted using 
        a Gaussian with modulation depending on the noise at the frequency $\omega_L$, 
				$g(t) = \text{exp}( -(t/T_2)^2 ) \text{exp}( -8 \sin^4(\omega_{\text{L}} t/4) 
				\gamma_\text{e}^2 \langle B_{\text{N}}^2 \rangle / \omega_{\text{L}}^2 )$ (solid orange line). 
        Here, $\gamma_{\text{e}}$ is the electron gyromagnetic ratio, 
				$\langle B_{\text{N}}^2 \rangle$ the root-mean-square of the noise field,
				and $\omega_{\text{L}}$ the proton Larmor frequency.
        This fit yields $T_2 = \SI{0.65(3)}{\micro\second}$.
		Panel (b): Measurement of $\langle S_i^y(t) S_i^y(0) \rangle$, the longitudinal autocorrelation
		in the doubly-rotating frame, under spin lock driving for a strong drive strength $\Omega_y^{\mathrm{S}}$ 
		to determine $T_{1\rho}$. The drive strength $\Omega_y^{\mathrm{S}}$ is chosen to be much larger than the on-site magnetic field strength due to the proton nuclear spin bath and the dipolar interaction strength of nearby surface spins.  This is modelled by an exponential function
        $\propto \text{exp}( -t/T_{1\rho} )$ obtaining $T_{1\rho} = \SI{26(3)}{\micro\second}$ (solid blue line).}
    \label{fig:spinDMFT_Experiment}
\end{figure}

\end{widetext}

In experiment, however, the autocorrelations behave very differently, as depicted in 
Fig.~\ref{fig:spinDMFT_Experiment}. In contrast to the spinDMFT result, the longitudinal decay is slowed down enormously, relative to the transversal one. Moreover, we find that the ratio between the decay times 
depends on the specific environment of the measured spin. Measurements at different spots on the diamond
surface lead to quantitatively different ratios. Yet the significant difference between
the longitudinal and the transversal channel is a common feature. 
As explained above, a variation of the ratio of the decay times
cannot be reproduced by spinDMFT because
the environment is described by a single environment field depending only on $J_{\text{Q,i}}$.
This caveat clearly calls for a methodological extension.

We stress that the key aspect of spinDMFT is to replace spin environments by dynamic random mean-fields
which is justified if the contributions from the 
environment are numerous and similar in magnitude. 
A formal expansion parameter is the inverse effective coordination number $z$ of a spin.
The term `effective' is used because not only the nearest neighbors matter for long-range
interactions. This effective coordination number
is given by the ratio of the linear sum squared and the quadratic sum of all couplings \cite{grass21},
\begin{align}
\label{eqn:coord-number}
    z_{i} &\eqqcolon \frac{\big(\sum_{j} |J_{ij}|\big)^2 }{\sum_{j} J_{ij}^2 }.
\end{align}
The larger $z$, the more the environment consists of many contributions justifying
the assumption that the environment behaves like a classical random variable
with normal distribution according to the central limit theorem.
For lattices, the coordination number is fairly large, e.g., $z \approx 19.1$ for the triangular lattice 
with couplings $J \propto 1/r^3$ \cite{grass21}. For inhomogeneous systems, this is not necessarily the case. Some strong constraints on the distances between spins
push the ratio up, but for  spins distributed totally at random 
one obtains values in the range $z \approx 1-10$ \cite{grass21}. 
Frequently, randomly positioned spins have only one or two close neighbors 
dominating their dynamics which limits the accuracy of the mean-field substitution of spinDMFT. 

For these reasons, an improved approach has to avoid replacing spin environments that correspond 
to small coordination numbers by classical fields. Since the experimentally measured 
autocorrelations vary from spin to spin, 
one has to incorporate the inhomogeneous nature of the system as well. These considerations
urge to develop a cluster mean-field approach as done in the next section.

\section{Cluster \enforce{spin}DMFT}
\label{sec:ClusterspinDMFT}

We argued that spinDMFT neglects spatial information to a large degree as is common
in mean-field theories which result in effective single-site problems by reducing
the environment to single fields. Thereby, subtle interference effects
of quantum processes are neglected. Indeed, spinDMFT is semiclassical in nature. The
local fields are taken to be classical so that quantumness is reduced to the quantum
effects of a single spin. Thus, our extension to clusters of spin will serve two purposes:
inclusion of spatial information and more accurate description of quantum dynamics.
As a consequence of dealing with clusters of spins on the quantum level, the mean-fields become 
weaker and the local dynamics more accurate. We call this extension cluster spinDMFT or CspinDMFT
for short. We derive it in the present section for an isotropic Heisenberg model
for simplicity.

\begin{figure}[htb]
    \includegraphics{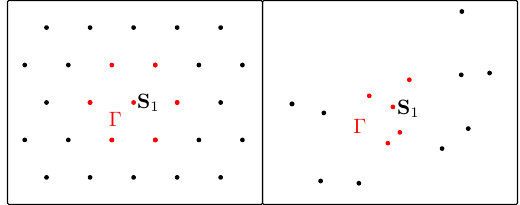}
    \caption{Examples for clusters $\Gamma$ around a central spin $\mb{S}_1$ for different graphs:
		the left panel displays the regular triangular lattice while the right panel a random seed.
		The couplings between the spins are considered to decrease with the distances with a power law.}
    \label{fig:examplecluster}
\end{figure}

We consider this model with $S~=~\tfrac12$ at infinite temperature of which the Hamiltonian reads
\begin{align}
    \mb{H} &= \frac12 \sum_{i\neq j} J_{ij}     \vec{\mb{S}}_i \cdot \vec{\mb{S}}_j.
    \label{eqn:ISO_Hamiltonian}
\end{align}
The underlying graph, i.e., the positions of the sites labeled $i$ and $j$, and the couplings $J_{ij}$ 
need not be specified for the time being. But we emphasize that the particular goal of the approach is to describe randomly distributed spins as in the experimental setup. The aim is to reliably compute the dynamics of a 
selected spin $\mb{S}_1$ which we call the central spin. To this end, we consider a group of spins in its proximity, see e.g.\ Fig.~\ref{fig:examplecluster}. This group of spins together with the central spin is henceforth called 
\emph{cluster}, denoted by the letter $\Gamma$.

For a cluster calculation, we have to define which spins around the central spin under study 
are chosen to form the considered cluster. Clearly, the spins which are coupled most strongly 
should be tracked.
But still there are (at least) two strategies to take the strength of the couplings into
account. Strategy (i) considers the coupling to the central spin to be decisive. Thus,
the decision which spin is included in the cluster is taken solely by its coupling
strength to the central spin. We call this strategy the \emph{central-spin-based strategy}.
It is certainly plausible, but it has a certain drawback: there can be a
spin $j$, strongly coupled to the central spin 1, which itself is strongly coupled to a spin
$m$. The latter, however, might be only weakly coupled to the central spin. Thus,
it would not be included in the cluster although it is obvious that it has a sizable
effect on the spin dynamics.

Thus, we also consider a recursive strategy (ii) which we call \emph{cluster-based strategy}
in which the spins are added one by one.
The decision which spin $m$ is to be included next in the cluster is taken based on
the total strength of \emph{all} its couplings to the spins already included in the cluster.
The details of both strategies are presented and discussed
in Appendix \ref{app:clusterchoice} and their performance is
discussed in Appendix \ref{app:convergence}. The following general procedure applies
to cluster from any conceivable strategy determining cluster. 
For the actual simulations below, we will use the second, cluster-based strategy.

In order to expose the details of CspinDMFT, it is helpful to analyze the different kinds of 
couplings in \eqref{eqn:ISO_Hamiltonian} with regard to $\Gamma$. We split 
the Hamiltonian as follows
\begin{subequations}
\begin{align}
    \mb{H} &= \frac12 \sum_{i\neq j | i,j \in \Gamma} J_{ij} \vec{\mb{S}}_i \cdot \vec{\mb{S}}_j \\ 
    &\quad + \sum_{i\in \Gamma} \vec{\mb{S}}_i \cdot \sum_{j\notin \Gamma} J_{ij} \vec{\mb{S}}_j \\
    &\quad + \frac12 \sum_{i\neq j | i,j \notin \Gamma} J_{ij} \vec{\mb{S}}_i \cdot \vec{\mb{S}}_j.
    \label{eqn:DRF_Hamiltonian2}
\end{align}
\end{subequations}
The first term contains the intra-cluster couplings; they are treated exactly in  CspinDMFT. 
The second term contains the couplings between spins of the cluster and the spins in the environment.
These couplings are treated on a mean-field level. The third term contains only spins of the environment
 which are decoupled from $\Gamma$. They do not show up in CspinDMFT.
The mean-field approximation consists in replacing the operators of the quantum environment 
\begin{align}
    \vec{\mb{V}}_{i} &\eqqcolon \sum_{j\notin \Gamma} J_{ij} \vec{\mb{S}}_j
\end{align}
by classical fields. In contrast to spinDMFT, there are several 
classical fields involved distinguished by the subscript $i$ because each spin 
in the cluster has its own field. Thus, an $N$-site cluster requires to track
$N$ classical fields which will be taken to be random in time and drawn from 
normal distributions.

Based on these considerations, CspinDMFT   works as follows
\begin{enumerate}
    \item[A] The quantum environment of each spin of the considered cluster is replaced
		by a time-dependent random classical mean-field drawn from a normal distribution.
		\item[B]The first moments of the distributions are set to zero.
    \item[C] The quantum mechanical expectation values of \mbox{$\mb{V}_i^\alpha$-$\mb{V}_j^\beta$} 
		correlations define the second moments of the normal distributions yielding self-consistency conditions 
		($\alpha,\beta\in\{x,y,z\}$).
    \item[D] The self-consistency conditions are solved iteratively.
\end{enumerate}

The substitution in step A is analogous to what is done in spinDMFT. 
The justifications are given in  Ref.\ \onlinecite{grass21}. Here, we simply substitute
\begin{align}
    \vec{\mb{V}}_{i} \to \vec{V}_{i}(t)
\end{align}
obtaining the mean-field Hamiltonian 
\begin{align}
    \mb{H}_{\Gamma}^{\text{mf}}(t) &= 
		\sum_{i \neq j | i,j \in \Gamma} J_{ij} \vec{\mb{S}}_{i} \cdot \vec{\mb{S}}_{j}
    + \sum_{i \in \Gamma} \vec{\mb{S}}_{i} \cdot \vec{V}_{i}(t).
\end{align}
The expectation values are computed by the trace over the density matrix at infinite temperature 
of the Hilbert space of the cluster combined with a classical average over the normal distributions of the 
 mean-fields \cite{grass21}. The required ingredient for the approach are the first and second moments 
of the mean-fields because they define the normal distributions.

Step C consists in relating the required moments to quantum expectation values of the corresponding environment operators. The first moments vanish due to the assumed infinite temperature
\begin{align}
    \overline{ V^{\alpha}_{i}(t) } &= \langle \mb{V}^{\alpha}_{i}(t) \rangle = 0.
\end{align}
For the second moments, we find
\begin{subequations}
\label{eqn:rawselfcons}
\begin{align}
    \overline{ V^{\alpha}_{i}(t) V^{\beta}_{j}(0) } &= \langle \mb{V}^{\alpha}_{i}(t) \mb{V}^{\beta}_{j}(0) \rangle \\
    &= \sum_{p,q \notin \Gamma} J_{ip} J_{jq} \langle \mb{S}^{\alpha}_{p}(t) \mb{S}^{\beta}_{q}(0) \rangle.
		\label{eqn:rawselfcons-b}
\end{align}
\end{subequations}
Since the spins in these expectation values are not in the cluster $\Gamma$ these moments cannot be
taken from the cluster calculation itself. In spinDMFT, this issue was circumvented by the following two 
ideas.

Any pair correlation, i.e., any correlation between two different spins, was neglected because
they are suppressed by the inverse coordination number. This step was mandatory in spinDMFT because the approach does not allow for the computation of pair correlations since the effective single site problem
comprises only a single spin. Obviously, CspinDMFT provides room for improvement on this aspect.

Second, in spinDMFT all autocorrelations are replaced by the autocorrelation of the considered central spin.
This results from the assumption that all spins of the system are essentially equivalent. While this assumption
holds in Bravais lattices, it must be questioned for inhomogeneous systems. 

\begin{figure}[htb]
    \includegraphics{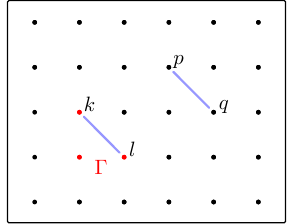}
    \caption{Example for a correlation replica on a square lattice. Translational invariance
		imposes that the pair correlation between the spins $p$ and $q$ is identical 
		to the pair correlation between the spins $k$ and $l$ so that they can be mutually substituted.}
    \label{fig:examplecorrelationreplica}
\end{figure}

Returning to Eq.\ \eqref{eqn:rawselfcons} let us replace the required, but unknown
correlations by some known correlations computed within the cluster similar to what is done in spinDMFT
for the autocorrelation. This idea can be implemented in many different ways. For lattices an obvious approach 
consists in exploiting the translational invariance and to substitute out-of-cluster correlations by their 
exact replicas in $\Gamma$ as shown exemplarily in Fig.~\ref{fig:examplecorrelationreplica}, see also 
Appendix \ref{app:ordered_system} for details.
For inhomogeneous systems, we pursue the analogous strategy by  identifying 
approximate replicas for each out-of-cluster correlation. We call this 
\emph{correlation replica approximation} (CRA) and provide its details next.

The goal of the CRA is to systematically find a representative correlation in $\Gamma$ for each correlation in 
$\overline{\Gamma}$, where $\overline{\Gamma}$ is the complement of $\Gamma$, i.e., all spins
that are not part of the cluster $\Gamma$. In mathematical terms, we look for  a map 
\begin{align}
    f: \overline{\Gamma}^2 \to \Gamma^2,
\end{align}
which maps index pairs $\{pq\}$ to $\{kl\}$ such that 
$\langle \mb{S}^{\alpha}_{p}(t) \mb{S}^{\beta}_{q}(0) \rangle$ and 
$\langle \mb{S}^{\alpha}_{k}(t) \mb{S}^{\beta}_{l}(0) \rangle$ are as similar as possible. 
Since none of the correlations is known a priori, defining such a map, which ensures the similarity
of the correlations, is non-trivial. 
One way to tackle this issue is to resort to the couplings and the short-time behavior
as a measure for similarity. In Appendix \ref{app:correlationassignment}, we analytically 
derive quantities suitable to measure similarity based on the correlation behavior around $t=0$. 
Eventually, we use the following map
\begin{subequations}
\begin{align}
    f(\{pq\}) &= 
    \begin{cases}
        \min_{kl \in \Gamma} \lvert J_{pq}^2 - J_{kl}^2 \rvert, & p \neq q, \\
        \min_{k \in \Gamma}  \lvert s_{p} - s_{k}  \rvert, & p = q, \quad \text{with}
    \end{cases} 
		\label{eqn:CRAmap}
		\\
    s_{p} &\coloneqq \sum_{r\neq p} J_{rp}^2.
    \label{eqn:quad_sum}
		\end{align}
\end{subequations}
Strictly, autocorrelations are mapped to autocorrelations and  pair correlations are 
mapped to pair correlations. In addition, it makes sense to define a lower cutoff for $J_{pq}^2$ by
\begin{align}
    c_{\text{cut}} &\eqqcolon \frac12 \min_{uv \in \Gamma} J_{uv}^2
\end{align}
such that any pair correlation in $\overline{\Gamma}$ with $J_{pq}^2 < c_{\text{cut}}$ is
considered too small to be relevant and hence is set to zero. This cutoff prevents that the weakest pair correlation in $\Gamma$ is overly weighted in
the self-consistency.

Using the map defined above, we can approximate the second moments of the mean fields
according to
\begin{align}
    \overline{ V^{\alpha}_{i}(t) V^{\beta}_{j}(0) } 
    &\approx \sum_{k,l \in \Gamma} \left(J^{\text{CRA}}_{ij,kl}\right)^2   
		\langle \mb{S}^{\alpha}_{k}(t) \mb{S}^{\beta}_{l}(0) \rangle,
    \label{eqn:scproblemtoymodel}
\end{align}
where  the coupling tensor $J^{\text{CRA}}_{ij,kl}$ is defined by
\begin{align}
    \left(J^{\text{CRA}}_{ij,kl}\right)^2 &\eqqcolon \sum_{\substack{ p,q \notin \Gamma, \\ 
		f(\{pq\}) = \{kl\}, \\ J_{pq}^2>c_{\text{cut}} }} J_{ip} J_{jq}
    \label{eqn:couplingtensor}
\end{align}
for all $i,j,k,l \in \Gamma$. The sum runs over all index pairs $\{pq\}$ that fulfill the constraints 
$f(\{pq\}) = \{kl\}$ and $J^2_{pq}>c_{\text{cut}}$. 

In this way, we obtain the closure of the self-consistency by 
Eq.~\eqref{eqn:scproblemtoymodel}.  We solve it numerically by iteration once
 the coupling tensor $J^{\text{CRA}}_{ij,kl}$ is known. For regular lattice problems 
the coupling tensor $J^{\text{CRav}}_{ij,kl}$ is used instead; it is defined in 
Eq.~\eqref{eqn:JCRA_ordered} in Appendix \ref{app:ordered_system}.

Technical details of the numerical implementation are provided in Appendix \ref{app:numerical_details}. 
In essence, the procedure is the same as for  spinDMFT \cite{grass21}. In Appendix \ref{app:convergence}, we 
illustrate that the results converge upon increasing the  cluster size. The following section is devoted to the application of the introduced method to the experimental scenario of dipolar surface spins in 
a doubly-rotating frame.

\section{Ensemble of dipolar surface spins}
\label{sec:ApplicationClusterspinDMFT}

Here we adapt the developed CspinDMFT to the experimental
setup. We adopt the Hamiltonian \eqref{eqn:DRF_Hamiltonian} which 
fixes a large number of parameters. The remaining parameters are the positions
of the spins and the global energy scale. For these, we will use the available input
from experiment.

As far as the CspinDMFT is concerned we follow the steps exposed in the previous section.
This leads us to the closed set of self-consistency conditions
\begin{align}
    \overline{ V^{\alpha}_{i}(t) V^{\beta}_{j}(0) } 
    &= D^{\alpha\alpha} D^{\beta\beta} \sum_{k,l \in \Gamma} 
		\left(J^{\text{CRA}}_{ij,kl}\right)^2 \langle \mb{S}^{\alpha}_{k}(t) \mb{S}^{\beta}_{l}(0) \rangle.
    \label{eqn:scproblemdipolemodel}
\end{align}
They are identical to the ones in Eq.\ \eqref{eqn:scproblemtoymodel} except for the 
anisotropic prefactors $D^{\alpha\beta}$ from Eq.~\eqref{eqn:DRF_localenvironments}. 
Employing the symmetry relations \eqref{eqn:symmetrierelations}
and inserting the prefactors, we finally arrive at 
\begin{subequations}
\begin{align}
    \overline{ V^{x}_{i}(t) V^{x}_{j}(0) } &= 
    \tfrac1{16} \sum_{k,l \in \Gamma} \left(J^{\text{CRA}}_{ij,kl}\right)^2 
		\langle \mb{S}^{x}_{k}(t) \mb{S}^{x}_{l}(0) \rangle \\
    \overline{ V^{y}_{i}(t) V^{y}_{j}(0) } &= 
    \tfrac1{4} \sum_{k,l \in \Gamma} \left(J^{\text{CRA}}_{ij,kl}\right)^2 
		\langle \mb{S}^{y}_{k}(t) \mb{S}^{y}_{l}(0) \rangle \\
    \overline{ V^{z}_{i}(t) V^{z}_{j}(0) } &= 
    \tfrac1{16} \sum_{k,l \in \Gamma} \left(J^{\text{CRA}}_{ij,kl}\right)^2 
		\langle \mb{S}^{z}_{k}(t) \mb{S}^{z}_{l}(0) \rangle.
    \label{eqn:scproblemdipolemodel2}
\end{align}
\end{subequations}
The coupling tensor $J^{\text{CRA}}_{ij,kl}$ is computed from the spin positions and the dipolar coupling in the doubly-rotating frame given in Eq.~\eqref{eqn:magic_angle_coupling}. 

\begin{figure}[htb]
    \includegraphics[width=0.5\textwidth]{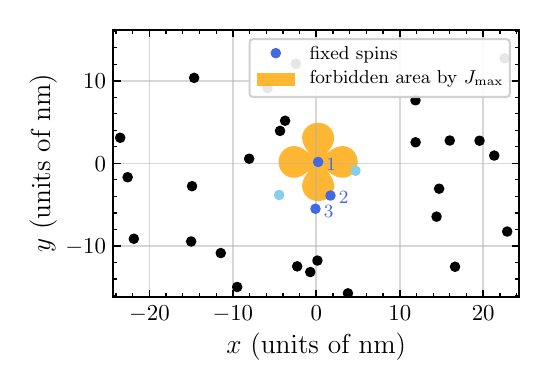}
    \caption{Cutout of a typical distribution of spins on the surface with the central spin being
		No.\ 1 in Fig.~\ref{fig:Application_MRandom_DDRF}. The dark blue dots correspond to the 
		spin positions measured in experiment. The positions of the other spins are drawn randomly 
		considering the constraints 3)-4).
		The two light blue spins are important for the dynamics of the central spin and, thus, are added to 
		the cluster in the CspinDMFT simulation. The remaining spins (black) form the mean fields. 
		We set the radii of the dots to $\SI{0.5}{\nano\meter}$ visualizing the constraint of the 
		minimum distance 3): two dots are not allowed to overlap. 
		Moreover, no spin can be in the orange area because otherwise another value of  $J_{\mathrm{max}}$
		would arise.}
    \label{fig:Application_example_picture} 
\end{figure}

In contrast to what was done in the previous section for inhomogeneous spin ensembles, 
the spin positions are not drawn completely at random, but there are some constraints known from the 
experiment
\begin{enumerate}
    \item[1)] Using methodology similar to the one employed in Ref.\ \onlinecite{sushk14}, we extract the positions of the central spin and its two strongest coupled neighbors from the measurements obtaining
    \begin{subequations}
    \begin{align}
        \vec{r}_1/\si{\nano\meter} &\eqqcolon \begin{pmatrix} 0.22 & ,& 0.17 \\ \end{pmatrix}^{\top}, \quad \text{(central spin)} \\
        \vec{r}_2/\si{\nano\meter} &\eqqcolon \begin{pmatrix} 1.7  & ,& -3.9 \\ \end{pmatrix}^{\top}, \\
        \vec{r}_3/\si{\nano\meter} &\eqqcolon \begin{pmatrix} -0.1 & ,& -5.5 \\ \end{pmatrix}^{\top}.
    \end{align}
    \label{eqn:fixedspinpos}
    \end{subequations}
    \item[2)] The modulus of the largest coupling to the central spin is 
    \begin{align}
        |J_{\text{max}}| &= |J(\vec{r_1} - \vec{r_2})| \approx \SI{3.078}{\mega\hertz}
    \end{align}
    Considering the couplings defined by Eq.~\eqref{eqn:magic_angle_coupling}, 
		this defines a glover-shaped area around the central spin, in which no other spin can be
		located because otherwise this spin would have the largest coupling to the central spin.
		This is illustrated in Fig.~\ref{fig:Application_example_picture}.
    \item[3)] Two spins are not allowed to have a  distance smaller than 
    \begin{align}
        d_{\text{min}} &\approx \SI{1}{\nano\meter} ,
    \end{align}
		i.e., we assign a certain spatial extent to the surface spin wave function.
    \item[4)] The spin density is given by \cite{rezaiarxiv}
    \begin{align}
        n_0 &\approx \frac1{64} \si{\nano\meter\tothe{-2}}.
    \end{align}
\end{enumerate}

Apart from these constraints, the spin positions are assumed to be random. In this way, we generate 
distributions such as the one in Fig.~\ref{fig:Application_example_picture}. 
In practice, we consider a square area with edge length $l$ and the three known spin sites from 
1). Subsequently, we successively draw $N-3$ random positions 
in the area fulfilling the constraint on the density $N/l^2 = n_0$ as well as the constraints 
2) and 3). For the couplings, we assume 
periodic boundary conditions: the shortest vector connecting two spins  defines their distance 
and thereby their coupling. This vector may cross the edges of the square. Typically, we take 
$l \approx \SI{200}{\nano\meter}$ because the coupling tensor $J^{\text{CRA}}_{ij,kl}$ is sufficiently 
converged for this size, i.e., it does not change significantly any more if $l$ is increased.
We stress that the computation of $J^{\text{CRA}}_{ij,kl}$ is extremely quick. The scaling of the computation is $\mathcal{O} ( N_{\Gamma}^4 (N-N_{\Gamma})^2 )$, but this is done before solving the self-consistency and, therefore, neither affected by the Monte-Carlo averaging nor by the time discretization. Thus, it is by no means the computational bottleneck so that $l$ could also be chosen larger.

Next, we define a cluster $\Gamma$ around the central spin. 
According to the conclusion in Appendix~\ref{app:convergence}, we use the cluster-based approach 
derived in \ref{subsec:cbstrategy} with cluster sizes $N_{\Gamma} = 5$ to $7$. 
Which value we actually choose is decided by comparing the strongest bond
between in-cluster spins and out-of-cluster spins
\begin{align}
    J_{\Gamma,\text{strongest}} &\eqqcolon \max_{k\notin \Gamma} \sum_{i \in \Gamma} |J_{ki}|.
\end{align}
This bond strength is a measure for the importance of a single spin outside of the cluster
on the spin dynamics within the cluster, see in Appendix~\ref{app:convergence}. 
Hence, the smaller $J_{\Gamma,\text{strongest}}$ is, the better.
We adopt the cluster size with the lowest value of $J_{\Gamma,\text{strongest}}$ because we saw
that cutting strong bonds by the choice of cluster deteriorates the results.
Subsequently, we compute the corresponding coupling tensor $J^{\text{CRA}}_{ij,kl}$. 
Once this is done, all ingredients for CspinDMFT are available and 
the self-consistency problem is solved numerically by iteration.

\begin{widetext}

\begin{figure}[b]
    \includegraphics[width=\textwidth]{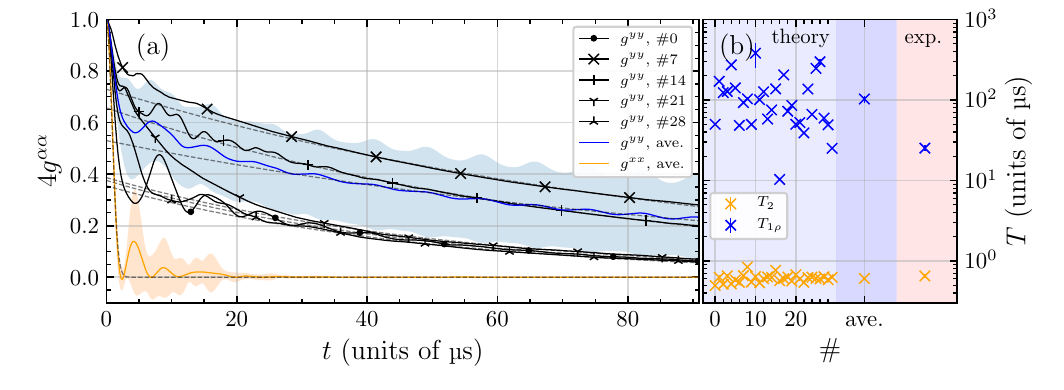}
    \caption{(a) Simulated longitudinal spin autocorrelations for the experimental setup of 
		three fixed spins and constrained random positions for the other spins. The average over $30$
		random sets of positions is also displayed as well as the corresponding $1\sigma$ interval of 
		the transversal and longitudinal autocorrelation, respectively. The tails of the 
		longitudinal curves ($t>\SI{30}{\micro\second}$) are fitted by exponentials 
		$\propto \mathrm{exp}(-t/T_{1\rho})$
		and the first drop of the transversal curves by Gaussians $g(t) = \mathrm{exp}( -(t/2T_{2})^2 )$. 
		The factor of 2 is required, since the $T_2$-measurement in experiment are performed in the 
		singly-rotating frame in which the couplings are twice as strong, see Eq.~\eqref{eqn:SRF_Hamiltonian}.
		The gray dashed lines render the fitting curves. (b) Characterstic times, i.e., the fit parameters,
		are shown in the light-bluish region for the various samples. The averaged values are shown in
		the dark-bluish region while the experimental values are shown in the reddish region.
		We emphasize the large ratio between the characteristic times in the longitudinal and the transversal channel.}
    \label{fig:Application_MRandom_DDRF} 
\end{figure}

\end{widetext}

In Fig.~\ref{fig:Application_MRandom_DDRF}(a), we present the results for several sets of random spin positions, while the positions of the three spins 1, 2, 3 are fixed as in Eq.~\eqref{eqn:fixedspinpos}. 
The common feature of the longitudinal autocorrelations is a quick initial drop to some moderate value between
$0.4$ and $0.8$  followed by a rather slow decay which can be well described by an exponential. 
The behavior of the transversal autocorrelations is qualitatively different.
They display a quick drop to zero which is sometimes followed by one or two small revivals. 
Essentially, the transversal signal has disappeared after about $\SI{10}{\micro\second}$, i.e., on a much shorter 
time scale than the longitudinal signal. 
Furthermore, it is striking that the first drop of the longitudinal signal varies relatively strongly from configuration to configuration of spin positions. This can also be seen in the relatively large $1\sigma$-interval of the averaged longitudinal signal. We conclude that the environment of the central spin 
even beyond the two closest spins, of which the 
positions are measured and thus given, matters considerably.

\begin{figure}[htb]
    \includegraphics[width=\columnwidth]{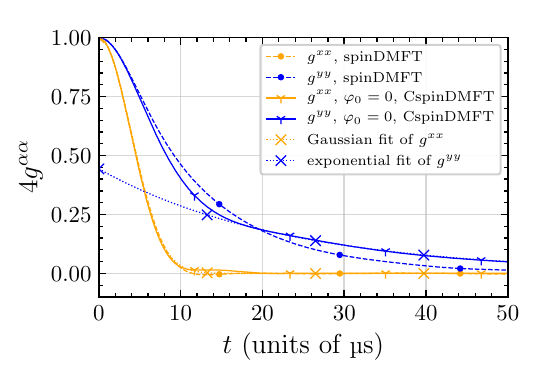}
    \caption{Spin-spin autocorrelations for dipolar couplings in the doubly-rotating frame
		as in Eq.\ \eqref{eqn:DRF_Hamiltonian} on a triangular lattice obtained from CspinDMFT 
		with $N_\Gamma = 5$. The lattice can be tilted relative to the reference axis for the couplings.
		Only the results for $\varphi_0=0$ are shown because the tilting angle has only a minor influence.
	  The longitudinal result ($t>\SI{20}{\micro\second}$) is fitted by an exponential 
		$\propto \mathrm{exp}( -t/T_{1\rho})$ to determine $T_{1\rho}$; 
		the transversal results by a Gaussian $g(t) = \mathrm{exp}(-(t/2T_{2})^2)$. The fits are 
		shown by dotted lines.}
    \label{fig:Application_Triangular_DDRF} 
\end{figure}

From experiment, we take that the characteristic transversal decay time $T_2$ 
is essentially determined by $1/J_\text{max}$ which is in-line with the
fact that this decay happens fast. In order to determine $T_2$ systematically
we fit the initial drop of the transversal signal by a Gaussian.
This appears to be appropriate for the numerical results obtained by CspinDMFT, 
 see for instance the orange curve in Fig.\ \ref{fig:Application_Triangular_DDRF}.
As displayed in Fig.~\ref{fig:Application_MRandom_DDRF}(b), 
the resulting times $T_2$ vary between $0.5-1.0\,\si{\micro\second}$ which is convincingly 
close to the experimental values. 

To extract $T_{1\rho}$ we use an exponential fit of the longitudinal 
autocorrelations after some offset in time because the initial drop and its subsequent oscillations
show a qualitatively different temporal behavior. The tails of the longitudinal autocorrelations, however,
are very well captured by exponential fits. We obtain, that the start time of each fit hardly matters, see Fig.~\ref{fig:Application_MRandom_DDRF}(a).
Yet we observe that the values for $T_{1\rho}$ vary considerably from configuration to configuration of random
spin positions. For the averaged signal, we obtain about $\SI{100}{\micro\second}$, 
but the characteristic times of single autocorrelations cover the range  $10-400\,\si{\micro\second}$. 
This observation underlines our above conclusion that the environment
of the central spin matters strongly. Even if the positions of two close spins are fixed the
behavior changes a lot from random set to random set of spin positions. 
This shows that the system is still far away from any self-averaging.

It is also worth mentioning that the variance of $T_{1\rho}$ becomes even larger when the positions of the closest neighbors of the central spin 
are also varied in the simulations. This matches qualitatively with the experimental observation of different relaxation times at 
different spots on the diamond surface, i.e., different geometries. Extracting spin positions from the measurement, however, forms a difficult and cumbersome task
which is why we did not compare theory and experiment for other configurations.

The key observation of Fig.~\ref{fig:Application_MRandom_DDRF} is the large ratio between the characteristic time scales in the
longitudinal and the transversal channel. There is more than one order of magnitude difference
between these times in experiment, see reddish area of panel (b). The ratio in the averaged theoretical
calculations even exceeds two orders of magnitude. But the individual ratios of each random set of spin
positions strongly fluctuate between one to almost two orders of magnitude. Where
does this large ratio stem from? In the Hamiltonian \eqref{eqn:DRF_Hamiltonian} 
there is only a factor of 2 difference between the longitudinal and transversal couplings
of the spin components. We claim that it is the positional randomness which makes the difference.

To corroborate this hypothesis we present
autocorrelations obtained by CspinDMFT for the Hamiltonian \eqref{eqn:DRF_Hamiltonian} on a \emph{regular
triangular lattice}, not on a randomly chosen set of sites. Thus, the coupling tensor entering the self-consistency 
conditions in Eq.~\eqref{eqn:scproblemdipolemodel} is changed and instead of using the CRA, we 
map correlations to their replicas based on translational invariance, see Appendix \ref{app:ordered_system}. 
Since the couplings depend on the angle $\varphi_{ij}$, see Eq.~\eqref{eqn:magic_angle_coupling},
the results will depend on how the reference axis for $\varphi$ is placed in the lattice. 
We take this into account by a tilting angle $\varphi_0$. It describes the angle between the axis 
set by $\varphi=0$ and one of the axes connecting two neighboring sites of the triangular lattice. 
The ratio of the longitudinal and transversal time scales 
only takes a moderate value 
\begin{align}
    \overline{\left(\frac{T_{1\rho}}{T_2}\right)} = \num{3.8(3)}, 
\end{align}
where the overbar stands for an average over the tilting angle, but all angles yield
essentially the same result. This is far too low in comparison to experiment.
This shows that a regular set of spin positions implies a spin dynamics which is at odds with
the experimental observations. We conclude that the randomness must be
a pivotal element in understanding the observed spin dynamics.

The fact that outliers of very large values of $T_{1\rho}$ occur in the random
ensembles is interesting in itself. It appears that
there exist configurations of spins which do not interact much with their environmental spins.
Hence, the eigenstates in these configurations must be well localized within the considered cluster.
One may speculate that the very slow decay of certain longitudinal autocorrelations is a precursor
of many-body localization. In the strict sense, true localization would imply that the autocorrelations
persist, i.e., the time scale would be infinite. On the one hand, the clusters we treat in CspinDMFT are
still fairly small with up to 7 spins for the results in Fig.\ \ref{fig:Application_MRandom_DDRF}.
Hence, it is possible that the decay is due to some `leakage' occurring due to the approximate treatment.
On the other hand, it is also conceivable that the non-infinite values of $T_{1\rho}$ reflect the
true physical behavior because recent results suggest that perfect many-body localization
is not generic \cite{evers23, sels23}.

The obtained agreement between experiment and theory is good and we conclude that
our approach provides a valid model for the experimental setup.  Yet the agreement is not 
quantitatively perfect. There are three main reasons for this. The first point is that 
our results clearly show that the randomness of the spin positions introduces
a large variability in the results. No self-averaging takes place. Furthermore,
it is not fully understood whether the spin positions change in the course of time
so that the experiment tends to measure averages \cite{dwyer22}. Generally, the lack of knowledge
on the whole spin systems also limits the accuracy of a theoretical description. 

Second, we introduced a powerful approximate approach by extending spinDMFT to CspinDMFT.
Yet it is still an approximation due to the finite cluster size
which can be simulated. Thus, a part of the discrepancy between theory
and experiment is likely to stem from the theoretical approximation.

Third, the theoretical description starts from the effective Hamiltonian \eqref{eqn:DRF_Hamiltonian}.
This is not the Hamiltonian realized directly on the diamond surface; rather it is
the effective Hamiltonian resulting from two nested rotating wave approximations in the
doubly-rotating frame. This must be realized in experiment and the measurements must
take these rotating frames into account. All imperfections in field alignment and
in pulse amplitude or duration will have an effect on the quantitative results.
For instance, the experimental data does not start precisely at $t=0$, but it has a certain intrinsic
offset explaining why the data points in Fig.\ \ref{fig:spinDMFT_Experiment} do not
start from 1 on the $y$-axis.

In view of these difficulties, the obtained agreement between experiment and theory 
is fully satisfactory.

\section{Conclusions}
\label{sec:Conclusion}

In this article, we pursued two objectives. The first is methodological, namely
the extension of the spin dynamic mean-field theory (spinDMFT) \cite{grass21} to
a cluster dynamic mean-field theory (CspinDMFT). The second is to understand the
experimental finding in  random dipolar spin ensembles that the longitudinal
relaxation is much slower than the transversal one.
In fact, the failure of spinDMFT to explain the experimental observation
triggered the strive for the methodological progress.

SpinDMFT  summarizes all couplings into one energy scale. This 
disregards any influence of the spatial positions of the spins.
CspinDMFT treats the dynamics in a cluster of spins rigorously and only the cluster's environment
is treated on a mean-field level. This improves the approximation because all processes
within the cluster are dealt with exactly. Thus, the whole spatial dependence within the cluster 
is captured: increasing the cluster size improves the accuracy.
We established systematic self-consistency
conditions reaching a closed set of equations for the random, normal distributed
mean fields. The convergence of CspinDMFT with increasing cluster size was found
to be excellent for a regular triangular lattice. It is also good for inhomogeneous spin ensembles
with randomness although the fluctuations in the spin dynamics
between different  sets of random positions is very high. The choice of appropriate clusters around 
the central spin needs careful consideration. One should avoid to ``cut'' strong couplings, i.e.,
to assign a spin to the cluster while another spin, to which it is strongly coupled, 
is incorporated in the mean fields. Strategies to find appropriate
clusters are suggested, tested, and discussed.

The developed CspinDMFT is applied to the effective Hamiltonian in the doubly-rotating
frame for a dipolar spin ensemble on the surface of diamonds at infinite temperature, i.e., for the completely disordered spin state. Experimentally, the
longitudinal and transversal autocorrelations are measured. A striking mismatch between
the relaxation times in these two channels is found. The ratio between the two
characteristic times is larger than one order of magnitude although there is only
a factor of 2 difference in the couplings. This finding is at odds
with what theory based on spinDMFT predicts so that the importance of the 
spin positions is underlined again. Similarly, CspinDMFT for a regular lattice shows a ratio
of less than 4 between the two characteristic times.
This disagrees with what is found experimentally for an ensemble with
random spin positions.

Hence, we are led to the conclusion that the randomness in the spin ensemble is the key ingredient
explaining the strongly differing time scales. Indeed, the theoretical predictions
by CspinDMFT for the ratio of time scales shows up to two orders of magnitude
difference. This agrees well with the experimental results
in view of the theoretical approximate treatment and the demanding experimental
realization of the effective Hamiltonian in the doubly-rotating frame. Scaling arguments suggest that the difference in the time scales becomes even larger, when the temperature is reduced \cite{nandk21}. 

Moreover, the very large ratio between the time scales in the two
channels found in CspinDMFT  can be interpreted as a precursor of 
 many-body localization, at least in the sense of very slow relaxation, but not of complete
persistence of correlations \cite{evers23}. Complete localization within the
studied cluster would imply the persistence of correlations, but it is
unlikely to occur in CspinDMFT because the approximation introduces 
 temporal randomness via the mean fields.

Our findings suggest a large variety of extensions. Clearly, further experimental
studies of dense spin ensembles are required to control and to understand
the relevant time scales in more detail. For instance, it would be extremely instructive to
examine spin ensembles in the whole range from perfect regularity, i.e., on lattices, 
to more and more irregular, random systems. This would allow one to make
further statements on the extent of slow relaxation and perspectively on many-body 
relaxation. 

From the theoretical side, many further applications of the 
developed mean-field approach are obvious. Clearly, a plethora of regular
and irregular geometries and spin-spin interactions can be tackled.
Moreover, one can also include external time dependencies. This will require
to refrain from using time translational invariance in the actual computations.
Similarly, one can envisage studying systems where the parameters such 
as the density of spins and their interactions vary in space. Combining
time and space dependence, phenomena of spin diffusion should be tractable.
An intriguing long-term vision consists in the consideration of spin systems at finite
temperature similar to previous studies \cite{gremp98,georg00a}.

    \begin{acknowledgments} 
        We thankfully acknowledge funding by the Deutsche Forschungsgemeinschaft (DFG) in projects
				UH90/13-1 and UH90/14-1 as well as the DFG and the Russian Foundation for Basic Research in TRR 160. 
				We also thank the TU Dortmund University for providing computing resources on the Linux HPC cluster (LiDO3), partially funded by the DFG in Project No. 271512359.
        The work at Boston University was supported by the US National Science Foundation Grant No. PHY-2014094.
    \end{acknowledgments}

\section{Experimental Considerations}
\label{app:experiment}

A near-surface nitrogen-vacancy (NV) center in diamond probes the dynamics of the surface spin system Fig.\ \ref{fig:SurfacePolarCoordinates}.
NV centers are optically initiated and read out by laser pulses delivered via a confocal microscopy setup.
Radiofrequency pulses that drive the NV center and surface spin transitions are delivered by a transmission line fabricated on a glass coverslip, on top of which the diamond sits.
A bias magnetic field of order 1000 G enables independent addressing of the NV spin and surface spin transitions by using different resonant RF tones, and all measurements take place at room temperature.

The dipolar coupling strength between the NV center and the surface spins is extracted from the double electron-electron resonance (DEER) measurement.
We limit this study to the case of one surface spin dominating the coupling with the NV center.  
In order to extract the positions of the three strongest coupled surface spins to the NV center, we perform DEER measurements at low external magnetic field, varying the azimuthal angle the external magnetic field makes with the NV axis.
This modulates the dipolar interaction strength between the NV center and each surface spin, and by using methods similar to those used in \onlinecite{sushk14}, we are able to extract  likely locations for these spins.

Each measurement presented in this work consists of a series of pulse sequences, which correlate the NV and central spin evolution, surrounded by NV optical spin polarization and readout steps.  
With this sequence, the NV center measures autocorrelation functions of the central surface spin $\langle S^{x,y,z}_i(t)S^{x,y,z}_i(0)\rangle$, where the time $t$ and the measured spin projection depend on the pulses applied during the evolution interval \cite{larao13}.

In this work, we consider measurements of the longitudinal and transverse spin autocorrelations within the doubly rotating frame. The transverse spin autocorrelation $\langle S^x_i(t)S^x_i(0)\rangle$ is extracted from a surface spin Hahn echo (T$_2$) measurement (Fig.\ \ref{fig:spinDMFT_Experiment}(a)), where the relaxation rate is directly related to the strength of the dipolar interaction of the central surface spin with nearby surface spins ($1/J_{\text{max}} = T_2$). In the doubly rotating frame, the longitudinal spin autocorrelation $\langle S^y_i(t)S^y_i(0)\rangle$ is extracted from a surface spin T$_{1\rho}$ measurement (Fig.\ \ref{fig:spinDMFT_Experiment} (b)), taken at a drive strength $\Omega_y^S$ much greater than the strength of on-site local magnetic fields.

\section{Strategies to define the cluster}
\label{app:clusterchoice}

In this appendix, we discuss strategies for choosing the cluster around the central spin 
in case of an inhomogeneous system of spins. The aim is to optimize the convergence of CspinDMFT by 
including the most important spins in the cluster so that their dynamics is treated exactly. 
Whether a spin is important or not is quantified by its contributions to the mean fields. 
We present two different strategies to pursue this goal. 
To be specific, we consider the isotropic Heisenberg model with $S=\tfrac12$. But the arguments
can straightforwardly be extended to more general couplings as well.

\subsection{Central-spin-based strategy}
\label{subsec:csbstrategy}

The first strategy is motivated by the general aim of CspinDMFT to find a good approximation for 
the dynamics of the spin under study, the so-called \emph{central} spin. 
Accordingly, it is plausible to treat the spins in the immediate vicinity exactly in the cluster
because they are most strongly coupled. 
To this end, we aim at making the mean field of the central spin as small as possible.
Since we do not want to run demanding numerical simulations \emph{before} the cluster
is defined we choose the initial moments as criterion for the size and importance of 
the mean field. Since the first moment vanishes due to the assumed initial disordered spin state 
reflecting infinite temperature, we consider the second moment. It is given by
\begin{subequations}
\begin{align}
    v_{1,\Gamma}^{\alpha}(0) &\eqqcolon \overline{ V_{1,\Gamma}^{\alpha}(0) V_{1,\Gamma}^{\alpha}(0) } \\
    &= \frac14 \sum_{p \notin \Gamma} J_{1p}^2 > 0,
\end{align}
\label{eqn:centralspinmeanfield}
\end{subequations}
where 1 is the index of the central spin and $\Gamma$ is the currently considered cluster. 
Next, we  extend the cluster by an additional spin with index $k$ according to 
\begin{align}
    \Gamma &\to \Gamma' = \Gamma \cup \{k\}.
\end{align}
The optimum choice of $k$ is the one that minimizes this second moment. 
Minimizing $v_{1,\Gamma'}^{\alpha}(0)$ with respect to $k$ corresponds to maximizing the difference
\begin{align}
    \Delta_{\Gamma\Gamma'}^{\alpha} &\eqqcolon v_{1,\Gamma}^{\alpha}(0) - v_{1,\Gamma'}^{\alpha}(0) 
		= \frac14 J_{1k}^2 > 0.
\end{align}
Hence, the spins to be added to the cluster should be determined from searching the 
maximum modulus of couplings to the central spin. This does not need to be done
iteratively, but one can simply choose the $N-1$ spins which are most strongly coupled
to the central spin. This is the central-spin-based strategy.

\subsection{Cluster-based strategy}
\label{subsec:cbstrategy}

An alternative approach aims at minimizing all mean fields of the cluster simultaneously. 
To this end, we consider the sum of the second moments
\begin{align}
    v_{\Gamma}^{\alpha}(0) &\eqqcolon \sum_{i,j \in \Gamma} 
		\overline{ V_{i,\Gamma}^{\alpha}(0) V_{j,\Gamma}^{\alpha}(0) } \\
    &= \frac14 \sum_{i,j \in \Gamma} \sum_{p,q \notin \Gamma} J_{ip} J_{jp}.
\end{align}
If the couplings are positive, minimizing $v_{\Gamma'}^{\alpha}(0)$ with respect to the added spin $k$ means maximizing the difference 
\begin{subequations}
\begin{align}
    \Delta_{\Gamma\Gamma'}^{\alpha} &\coloneqq v_{\Gamma}^{\alpha}(0) - v_{\Gamma'}^{\alpha}(0) \\ 
    &= \frac14 \sum_{i,j\in\Gamma} J_{ik} J_{jk}
    = \left( \frac12 \sum_{i\in\Gamma} J_{ik} \right)^2 > 0.
\end{align}
\end{subequations}
Thus, the next spin to be added can be determined from searching the maximum 
of the modulus of the linear sum of the couplings to all spins in the cluster
$\Gamma$. In the experimental scenario in Sec.~\ref{sec:ApplicationClusterspinDMFT}, however, the couplings become negative for half of the orientations of the distance vector. Therefore, contributions of different spins sometimes cancel one another in the linear  sum of the couplings so that 
this sum is not a good measure for the actual ``importance'' of a spin to the cluster. 
For this reason, we recommend to maximize the linear sum of the \emph{moduli} of the couplings instead. From our experience, this results in more appropriate clusters.

The results stemming from the two strategies are shown and compared in Appendix 
\ref{app:convergence}. Altogether, we find slight advantages for the cluster-based strategy.

\section{Mapping of out-of-cluster correlations to in-cluster correlations}
\label{app:correlationassignment}

In this appendix, we discuss the identification of out-of-cluster correlations with
in-cluster correlations which is needed to close the self-consistency problem in 
Eq.~\eqref{eqn:rawselfcons}. 

\subsection{Inhomogeneous systems}
\label{app:inhomsystem}

For the inhomogeneous system, we introduce the correlation replica approximation CRA 
which approximates correlations outside the cluster by correlations inside the cluster.
For this purpose, the correlations are classified according to their short-time
behavior as derived here. We consider the general correlation
\begin{align}
    g^{\gamma\rho}_{pq}(t) &\eqqcolon \langle \mb{S}^{\gamma}_{p}(t) \mb{S}^{\rho}_{q}(0)\rangle,
\end{align}
in the isotropic Heisenberg model with $S=\tfrac12$  defined by the Hamiltonian
\begin{align}
    \operator{H} &= \frac12 \sum_{i\neq j, \alpha} J_{ij} \mb{S}_i^\alpha \mb{S}_j^\alpha,
    \label{eqn:ISO_Hamiltonian_for_CRA}
\end{align}
with arbitrary, but fixed couplings $J_{ij}$. 
At $t=0$, the correlation is given by
\begin{align}
    g^{\gamma\rho}_{pq}(0) &= \frac14 \delta_{\gamma\rho} \delta_{pq}.
\end{align}
This result already provides important information for finding correlation replicas: autocorrelations take their maximum value at $t=0$ given by $\frac14$ 
while pair correlations vanish. Therefore, the most important characteristic is whether $p$ and $q$ are equal or not. The first temporal derivative 
\begin{align}
    \frac{dg^{\gamma\rho}_{pq}}{dt}(0) &= 0
\end{align}
vanishes for all $p,q$ and does not provide helpful information. The second derivative reads
\begin{subequations}
\begin{align}
  \frac{d^2g^{\gamma\rho}_{pq}}{dt^2} (t)
	&= - \langle \left[ \mb{H}, \left[ \mb{H}, \mb{S}^{\gamma}_{p} \right] \right](t) \mb{S}^{\rho}_{q}(0)\rangle \\
    &= \langle \left[ \mb{H}, \mb{S}^{\gamma}_{p} \right](t) \left[ \mb{H},\mb{S}^{\rho}_{q}\right](0)\rangle.
		\label{eq:deriv1}
\end{align}
\end{subequations}
The commutator with the Hamiltonian yields 
\begin{subequations}
\begin{align}
    \left[ \mb{H}, \mb{S}^{\gamma}_{p} \right] &= \mathrm{i} \sum_{i, i\neq p} \sum_{\alpha\beta} J_{pi} \epsilon^{\alpha\gamma\beta} \mb{S}_{i}^{\alpha} \mb{S}_{p}^{\beta}
\end{align}
\end{subequations}
which we insert in \eqref{eq:deriv1} to obtain 
\begin{align}
      \frac{d^2g^{\gamma\rho}_{pq}}{dt^2} (0)
    &= -\frac{\delta_{\gamma\rho}}{8}   \delta_{pq} \sum_{q \neq p} J_{pq}^2  +\frac{\delta_{\gamma\rho}}{8}   ( 1 -\delta_{pq} ) J_{pq}^2.
\end{align}
The prefactor $\delta_{\gamma\rho}$ implies that the second derivative contributes only for $\gamma = \rho$. 
Here this is no caveat because any correlation with $\gamma \neq \rho$ vanishes 
in the isotropic Heisenberg model at infinite temperature. 
For other models, this point might be worth reconsidering. Here, we can simply reduce the crucial quantities to 
\begin{align}
    s_{p} &\eqqcolon \sum_{q \neq p} J_{pq}^2
\end{align}
for the autocorrelations and to $J_{pq}^2$  for pair correlations with $p\neq q$.
Thus, we assess the similarity of correlations on the basis of the similarity of $s_p$ for autocorrelations
and on the basis of the similarity of $J_{pq}^2$ for pair correlations.
We use these characteristics also for the anisotropic model  in Sect.~\ref{sec:ApplicationClusterspinDMFT}.

\subsection{Regular lattice systems}
\label{app:ordered_system}

For regular systems with translational invariance, the mapping of the correlations in 
Eq.~\eqref{eqn:rawselfcons} is much easier. We discuss a systematic mapping 
restricting ourselves to Bravais lattices. The extension to crystal lattice with several
atoms in the unit cell is straightforward.

Each site of the lattice is completely equivalent. Hence, all autocorrelations and all pair correlations with the same distance vector $\vec{r}_{pq}$ are equal.  CspinDMFT breaks the translational symmetry by singling
out the cluster. So the contribution of the mean fields varies from site to site in $\Gamma$ and so do the correlation functions. This leads to ambiguities in finding the correlation replicas: for instance, the autocorrelation in the lattice can be represented by $N_{\Gamma}$ different autocorrelations in $\Gamma$ 
which need not be exactly the same. Here $N_{\Gamma}$ is the number of sites in the cluster.

A good way to deal with the ambiguity is to approximate the unknown out-of-cluster correlation with
the distance vector $\vec{r}_{kl}$ by
the average of all in-cluster correlations with the same distance vectors
\begin{align}
    \langle \mb{S}^{\alpha}_{p}(t) \mb{S}^{\beta}_{q}(0) \rangle
    \approx \frac1{N_{kl}} \sum_{\substack{ k,l \in \Gamma, \\ \vec{r}_{kl} = \vec{r}_{pq} }} 
		\langle \mb{S}^{\alpha}_{k}(t) \mb{S}^{\beta}_{l}(0) \rangle.
\end{align}
The sum runs over all index pairs $k,l$ with $\vec{r}_{kl}~=~\vec{r}_{pq}$, that is, all $N_{kl}$ correlation replicas within the cluster $\Gamma$. Then, the coupling tensor in Eq.~\eqref{eqn:scproblemtoymodel} reads
\begin{align}
    \left(J^{\text{CRav}}_{ij,kl}\right)^2 &= \frac1{N_{kl}} \sum_{\substack{ p,q \notin \Gamma, \\ 
		\vec{r}_{pq} = \vec{r}_{kl} }} J_{ip} J_{jq},
    \label{eqn:JCRA_ordered}
\end{align}
where CRav stands for \emph{correlation replica average}. 

There are other options as well, for instance using the correlations involving the central spin 
because one may expect that the dynamics of this spin is captured best. Our checks revealed that
this alternative affects the spin dynamics hardly because (i) the correlations in $\Gamma$ with the same 
$\vec{r}_{kl}$ do not differ much and (ii) because the final results are not very  sensitive 
to small variations in the temporal behavior of the mean fields.

\section{Numerical Details}
\label{app:numerical_details}

Here we discuss several aspects of the numerical implementation and point out technical issues
and how they can be dealt with.

\subsection{Implementation and numerical resources}
\label{app:implementation}

The implementation of the self-consistency problem is close to that of spinDMFT which is why we refer to the corresponding section in Ref.~\cite{grass21}. Clearly, the numerical effort for CspinDMFT is significantly larger
in several steps. First, the dimension of the covariance matrix built from the second moments of the mean fields 
becomes larger by the factor $N_{\Gamma}$ because each site in the cluster has its own mean field.
Second, the computation of the quantum expectation values is more demanding since the dimension of the 
Hilbert space  is $(2S+1)^{N_{\Gamma}}$ in the cluster instead of only $2S+1$. Third, one has to 
compute $N_{\Gamma}^2$ times more spin-spin correlations in each iteration step to close the self-consistency. This factor can be reduced to $N_{\Gamma}(N_{\Gamma}+1)/2$ by making use of time-reversal invariance. 
Fortunately, we still observe a fast convergence of the self-consistency iteration for CspinDMFT within
 only 3-5 steps. For typical numerical parameters (200 time steps and $10^4$ mean-field samples), cluster sizes up to $N_{\Gamma}=6$ require moderate numerical resources, e.g., the run time is $\approx 50 \, \text{core hours}$.

Finally, we emphasize the strong advantage of employing commutator-free exponential time (CFET) propagation to
compute time evolution operators \cite{alver11a}. Especially in the considered inhomogeneous systems, spatially
close spins are strongly coupled and thus display fast oscillations. 
Counterintuitively, these oscillations need not be resolved by fine time steps 
when using CFET propagation because the dynamics induced by the time-independent part of the Hamiltonian is captured exactly. Only the oscillations in the time-dependent mean fields have to be resolved to prevent numerical errors 
stemming from too coarse time discretization. 

\subsection{Definiteness of the covariance matrix}
\label{app:definiteness}

The substitution of quantum operators by classical Gaussian variables entails a numerical subtlety. 
To be able to interpret the quantum expectation values as matrix elements of a covariance matrix,
symmetry and positive semi-definiteness are necessary. Due to the considered infinite temperature, 
symmetry 
\begin{align}
\langle \mb{V}_i^\alpha(t)\mb{V}^{\beta}_j(0)\rangle = \langle \mb{V}^{\beta}_j(0) \mb{V}_i^\alpha(t)\rangle
\end{align}
 is always ensured. Positive semi-definiteness can be verified
 by summing the mean-field correlations with arbitrary real-valued coefficients $\lambda$ over all occurring indices according to
\begin{align}
    \sum_{i,j,\alpha\beta,t,t'} \langle \mb{V}_i^\alpha(t)\mb{V}^{\beta}_j(t')\rangle 
		\lambda_{i, \alpha, t} \lambda_{j, \beta, t'} \stackrel{!}{\geq} 0 ,
    \label{eqn:positivity}
\end{align}
where the time is also discretized so that the sum is well-defined.
Re-writing this expression as 
\begin{align}
\begin{split}
    \sum_{i,j,\alpha\beta,t,t'} & \langle \mb{V}_i^\alpha(t)\mb{V}^{\beta}_j(t')\rangle \lambda_{i, \alpha, t} \lambda_{j, \beta, t'} 
		\\
    &= \Big\langle \Big( \sum_{i,\alpha,t} \mb{V}_i^\alpha(t) \lambda_{i, \alpha, t} \Big)^2 \Big\rangle \geq 0
\end{split}
\end{align}
its non-negativity is obvious. Note that the basis for this conclusion is that
the covariance is indeed a product of the two independent environment operators.

But employing an approximation to the right hand side of Eq.~\eqref{eqn:rawselfcons-b} 
such as the CRA leading to Eq.~\eqref{eqn:couplingtensor} or Eq.~\eqref{eqn:JCRA_ordered}
one realizes that this is no longer the case because coupling tensor $J^{\text{CRA}}_{ij,kl}$ 
cannot be split into the product of two independent sums over $k$ and $l$,
see Eq.~\eqref{eqn:scproblemtoymodel}. Thus, the non-negativity of the approximate
covariance matrix cannot and is not guaranteed.

\begin{figure}[htb]
    \includegraphics[width=0.5\textwidth]{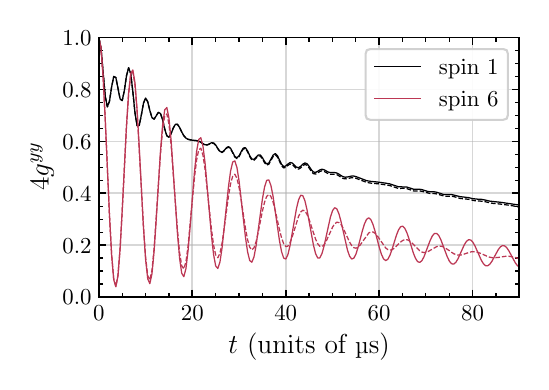}
    \caption{Exemplary results for a set of spin positions with noticeable violation of 
		the positive semi-definiteness of the approximated covariance matrix in the numerical simulation, 
		$r\approx 0.12$ for the effective dipolar Hamiltonian \eqref{eqn:DRF_Hamiltonian} 
		in the doubly-rotating frame. The cluster size 
		is $N_\Gamma = 6$.}
    \label{fig:definiteness}
\end{figure}

One way to quantify the violation of positive semi-definiteness due to the CRA
is to measure the ratio of the sum of all negative eigenvalues of the covariance matrix
over the sum of all positive eigenvalues $\mu_i$, i.e., 
\begin{align}
    r &= \frac{\sum_{\mu_i < 0} | \mu_i |}{\sum_{\mu_i > 0} \mu_i}.
\end{align}
For the eigenvalues of the matrix with matrix elements $\sum_{k,l \in \Gamma} \left(J^{\text{CRA}}_{ij,kl}\right)^2$
we find $r\approx 0.6$ which seems to indicate a strong violation of positive definiteness. 
But the relevant covariance matrix includes the spin-spin correlations, see 
Eq.~\eqref{eqn:scproblemtoymodel}. The  inclusion of these correlations 
reduces $r$ considerably. Thus, one can set the remaining negative eigenvalues to zero so that
the resulting matrix can be taken as covariance matrix of a normal distribution.

To assess the influence of this truncation in the numerical simulations, we track $r$
 in the final iteration step to check whether the truncation is justified and the obtained results are reliable. 
From our experience, the violations of positive semi-definiteness become relevant only in the inhomogeneous systems. Here, the violations are sometimes $r\gtrapprox 0.1$; we observed this in one out of 90 random sets of the spin positions.
For  all practical purposes, this is still tolerable. 
To show this, we exemplarily compare the results for a random set of positions 
with $r\approx 0.1$, where we (a) truncate all negative eigenvalues to zero
and (b) replace all negative eigenvalues by their absolute value. 
Fig.~\ref{fig:definiteness} depicts  both results and the differences are still extremely 
 small for the central spin (spin 1). For the outermost spin of the cluster (spin 6), 
the differences are larger. This is not surprising since the truncation affects the mean fields in the first place 
which are stronger for the spins at the border of the cluster. For violations far beyond 
$r\approx 0.1$ the obtained data might not be reliable, but  we did not observe such cases.

\subsection{Numerical error sources}
\label{app:stat_error}

Several numerical errors emerge in the numerical evaluation of the self-consistency problem. As in the original spinDMFT \cite{grass21}, we have a statistical error due to the Monte-Carlo sampling, an error from the time-discretization as well as a self-consistency error resulting from terminating after a small number of the iterations.

The statistical error of a spin correlation $g^{\alpha\beta}_{ij}(t)$ is given by
\begin{align}
    \sigma\left( g^{\alpha\beta}_{ij}(t) \right) &= 
    \frac1{\sqrt{M}} \sigma\left( g^{\alpha\beta}_{ij}\left(t,\vec{\mathcal{V}}\right) \right)
\end{align}
where $M$ is the sample size and $\vec{\mathcal{V}}$ stands for a single mean-field sample. We conservatively estimate the single-sample standard deviation by  
\begin{align}
    \sigma\left( g^{\alpha\beta}_{ij}\left(t,\vec{\mathcal{V}}\right) \right) &= \frac1{4 \sqrt{3}}
\end{align}
in accordance with the results in Ref.~\onlinecite{grass21}. Current numerical simulations indicate, that this value becomes smaller if the cluster size is increased. Harnessing this effect can considerably improve the performance.

The error from the time-discretization is difficult to analyze because it strongly depends on the considered system and geometry. To compute time-evolution operators, we use CFETs of order $2$ and integrate with the trapezoidal rule which typically makes the error scale as $\delta t^2$ where $\delta t$ is the (equidistant) step width. This scaling cannot be used for an estimate of the accuracy of the discretization. Hence, we choose the simple
strategy to compute the deviation between two different discretizations. If the deviation is not sufficiently small, the discretization has to be made finer.

Analogous to spinDMFT, the self-consistency problem is solved by iteration. We make some initial guess for the mean-field moments, compute the spin autocorrelations by Monte-Carlo simulation and, subsequently, update the mean-field moments via the self-consistent equations. This procedure is repeated until the results have sufficiently converged. One way to measure the convergence is to compute the time-averaged deviations between the current $(n)$ and the previous iteration step $(n-1)$ according to 
\begin{align}
    \Delta I_{ij}^{\alpha\beta}(n) &= \frac1{L+1} \sum_{l=0}^{L} | g^{\alpha\beta}_{ij,(n)} (l \delta t) - g^{\alpha\beta}_{ij,(n-1)} (l \delta t) | 
\end{align}
where $L$ is the number of time steps. The largest $\Delta I_{ij}^{\alpha\beta}(n)$ can then be compared to some preset tolerance to decide whether the iteration can be terminated. In this case, the iteration error can be estimated by the tolerance. 
The tolerance is chosen such that the iteration error does not exceed the statistical
error.

For best efficiency, the numerical parameters should be chosen such that all error sources are of the same magnitude. Throughout this article, the numerical error of any shown correlation is 
 $\SI{1}{\percent}$ or less of the theoretical maximum value $0.25$.

\section{Convergence of C\enforce{spin}DMFT}
\label{app:convergence}

Here we illustrate that the proposed extended mean-field approach converges
for increasing cluster size. This fact corroborates the justification of CspinDMFT.
We consider the regular triangular lattice with an isotropic spin Hamiltonian
first and, subsequently, inhomogeneous spin ensembles with random spin positions.

On the triangular lattice we choose couplings with a power-law dependence like dipolar
couplings
\begin{align}
    J_{ij} &= \frac{ r_0^3 \mathcal{J}_{0,\text{tri}}}{r_{ij}^3}  
		= \left( \frac{\sqrt{3}}{2} \right)^{\frac32} \frac{a^3\mathcal{J}_{0,\text{tri}}}{r_{ij}^3},    
\end{align}
where $r_0$ is defined via the spin density $n_0 = 1/r_0^2$ and $a$ is the lattice constant. 
The number $N_{\Gamma}$ of spins in the cluster is increased successively according to the numbering
 in panel (b) of Fig.~\ref{fig:results_triangular}. The energy constant $\mathcal{J}_{0,\text{tri}}$
sets the energy scale. For each $N_{\Gamma}$, we
compute the  coupling tensor \eqref{eqn:couplingtensor} and
 solve the self-consistent equations by iteration.
Due to the isotropy of the considered model, all off-diagonal correlations vanish while all diagonal correlations are equal
\begin{subequations}
\begin{align}
    \langle \mb{S}^{\alpha}_{k}(t) \mb{S}^{\beta}_{l}(0)\rangle &= 0 \qquad \forall \alpha \neq \beta, \\
    \langle \mb{S}^{x}_{k}(t) \mb{S}^{x}_{l}(0)\rangle &= \langle \mb{S}^{y}_{k}(t) \mb{S}^{y}_{l}(0)\rangle =
    \langle \mb{S}^{z}_{k}(t) \mb{S}^{z}_{l}(0)\rangle.
\end{align}
\label{eqn:symmetrierelations}
\end{subequations}

The obtained results for the autocorrelation of the central spin are shown in Fig.~\ref{fig:results_triangular} 
in panel (a) as function of time. Generally, the curves do not vary much with the cluster size. This is expected, since the effective coordination number of the triangular lattice $z$ defined in Eq.~\eqref{eqn:coord-number} 
 is fairly large, $z \approx 19.1$ \cite{grass21}, so that spinDMFT 
represents already a good approximation. At $t \approx 2/\mathcal{J}_0$, a low maximum appears which grows
upon increasing the cluster size, see also inset of panel (a). Its growth stops  at $N_{\Gamma} = 7$ 
indicating that it is important to include all nearest neighbors of the central spin in the cluster as
one may have expected a priori. The deviations of the results between $N_{\Gamma} = 7, 8$, and $9$ 
are very small and only slightly above the expected numerical errors resulting from the time discretization. 
Hence, at cluster sizes beyond the size of the cluster including the nearest neighbors, 
the advocated CspinDMFT constitutes a considerable improvement over spinDMFT for lattices. 
It converges well with the cluster size.

The main goal of this article is to access the spin dynamics in inhomogeneous systems.
As pointed out in the Introduction \ref{s:intro},  spinDMFT fails for inhomogeneous
system since it does not capture any aspect of the inhomogeneity beyond a global energy scale.
Moreover, the effective coordination number tends to be small in random systems.
Thus, we address the convergence of CspinDMFT for an inhomogeneous systems next.
We consider spins at randomly drawn positions in two dimensions. For simplicity, the spins are coupled 
according to the isotropic Heisenberg Hamiltonian \eqref{eqn:ISO_Hamiltonian} with 
\begin{align}
    J_{ij} &= \frac{r_0^3\mathcal{J}_{0,\text{inh}}}{r_{ij}^3},
\end{align}
where $r_0$ is defined via the spin density $n_0 = 1/r_0^2$.
It is expected that the environment and hence the spin dynamics varies strongly from spin to spin. 
As a consequence, the autocorrelations  depend strongly on the drawn set of spin positions. 
To capture this aspect, we study two different sets of spin positions: one where the central
spin is only weakly coupled to its environment and one where it is strongly coupled.
The results obtained by CspinDMFT are shown in Figs.~\ref{fig:weakISO} and~\ref{fig:strongISO}. 
We also compare the influence of the two strategies to define the cluster
introduced in Appendix \ref{app:clusterchoice}. 

Our first observation is that the correlations still deviate quite noticably when passing from two over three
to four spins in the cluster. Hence, one should not expect reliable  results for $N_{\Gamma}<4$. 
For $N_{\Gamma}\geq 4$, the general features of the curves, i.e., the time scale of the initial decay 
and the period of the dominant oscillation do no longer change in the displayed data. 
The remaining differences are small and concern only certain quantitative details such as the height 
of secondary extrema. Both the central-spin-based strategy (see upper rows of panels)
as well as the cluster-based strategy (see lower rows of panels) show sufficient 
convergence for $N_{\Gamma} \geq 4$. Inspecting the cluster shown  Fig.~\ref{fig:weakISO}(d), 
one can doubt the convergence of the cluster-based strategy
 because the cluster does not seem to be centered around the central spin. 
One spots at least three other close neighbors that are ignored in this choice of the cluster. 
But inspection of the results of the alternative strategy in panel (a) clarifies
that an explicit quantum-mechanical treatment of these spins (4, 5, and 6) 
has no large effect on the autocorrelation of the central spin. In addition, the computed
autocorrelations displayed in Fig.~\ref{fig:weakISO} (c) show  very good convergence
corroborating this choice of the cluster.

The convergence of the autocorrelation of the central spin with a stronger coupled environment
depicted in Fig.~\ref{fig:strongISO} is also good except for the outlier at $N_{\Gamma} = 8$ in panel (a). 
The occurrence of this outlier can be understood by inspecting the geometry and the clusters in panel (b). 
Spin 8 has an extremely close neighbor which is not included in the cluster. 
Including spin 8 in the cluster, but treating its strongly coupled neighbor only on the the mean-field 
level implies to cut a spin dimer which is obviously not justified. 

We conclude, that a good choice of the cluster needs to include at least $N_{\Gamma} = 4$ spins. 
It is advantageous if the couplings between in-cluster spins and out-of-cluster
spins are rather weak. It is difficult to provide compelling evidence to rank the two
strategies for the choice of cluster because of the strong fluctuations in inhomogeneous
systems. But the observation that cutting strong couplings between spins inside and outside
the cluster is detrimental leads us to conclude that the cluster-based strategy provides
better performing clusters than the central-spin-based strategy. 
For the decision to include or to exclude an additional spin in the cluster
the cluster-based strategy considers the couplings of this spin to all spin in the cluster 
so that ``cut'' dimers tend to be avoided. In contrast, the central-spin-based strategy is blind
to any couplings but to the ones to the central spin.

\begin{widetext}

\begin{figure}[htb]
    \includegraphics[width=\textwidth]{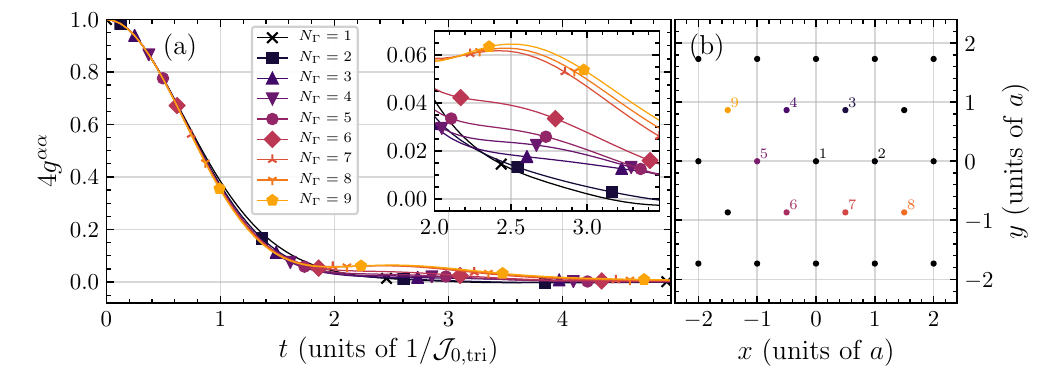}
    \caption{(a) Spin-spin autocorrelations in CspinDMFT on the triangular lattice at infinite temperature 
		for various cluster sizes $N_{\Gamma}$ according to the numbering in panel (b);
		$N_{\Gamma}=1$ refers to spinDMFT. Beyond the relatively large jump
		from $N_{\Gamma} = 6$ to $N_{\Gamma} = 7$ the deviations are barely visible.  
		(b) Triangular lattice and sequence of the spins included in the cluster $\Gamma$.}
    \label{fig:results_triangular}
\end{figure}
\begin{figure}[htb]
    \includegraphics[width=0.88\textwidth]{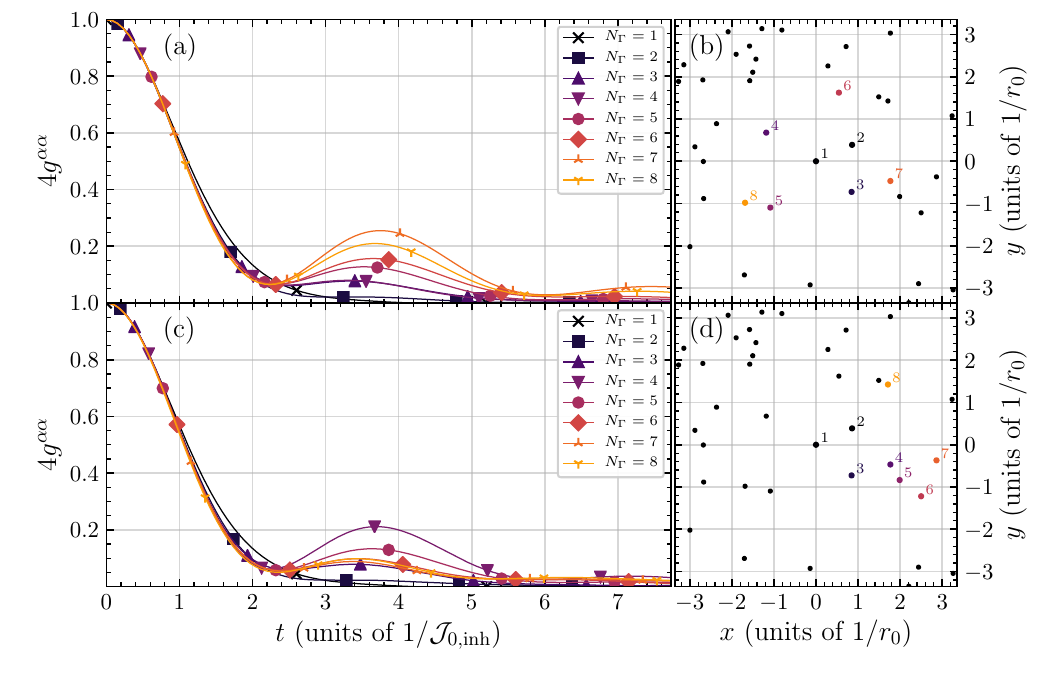}
    \caption{(a) Spin-spin autocorrelations in CspinDMFT of an inhomogeneous spin ensemble 
		with isotropic couplings at infinite temperature for various cluster sizes $N_{\Gamma}$
		according to the numbering in panel (b);	$N_{\Gamma}=1$ refers to spinDMFT. Here, a central 
		spin is considered which is coupled rather weakly to its environment. 
		The cluster was determined using the central-spin-based strategy \ref{subsec:csbstrategy}. (b) Spin positions
		and numbering of the spins in the considered clusters. (c) Same as in (a) using the 
		cluster-based strategy \ref{subsec:cbstrategy}. (d) Same as (b) based on  the 
		cluster-based strategy \ref{subsec:cbstrategy}.}
    \label{fig:weakISO}
\end{figure}
\begin{figure}[htb]
    \includegraphics[width=0.88\textwidth]{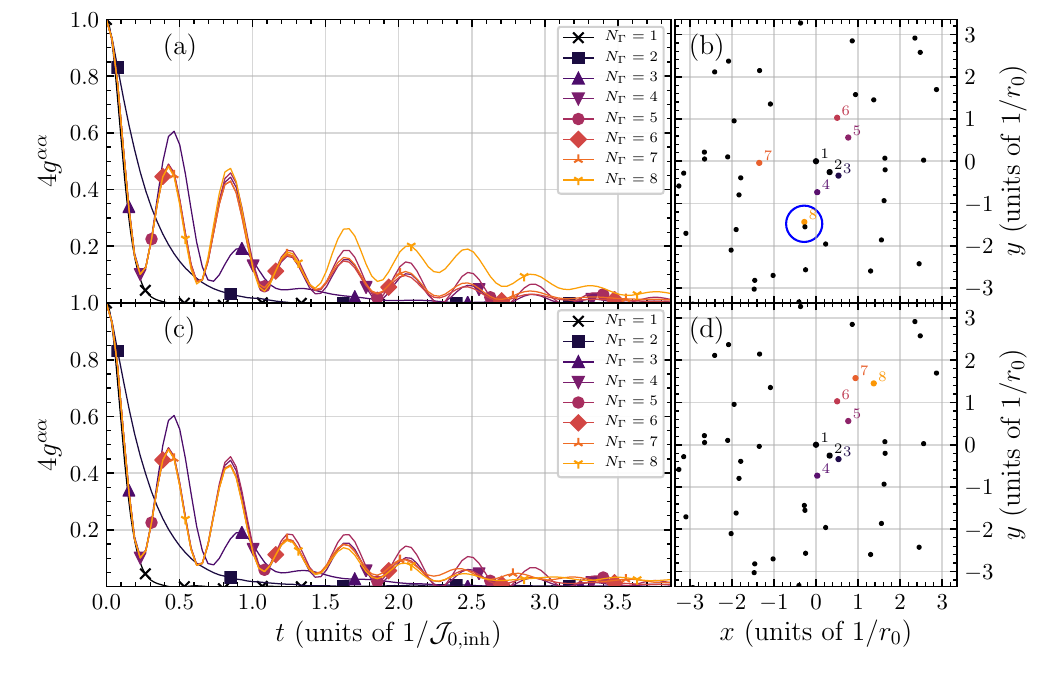}
    \caption{Same as in Fig.~\ref{fig:weakISO}, but here for a central spin which is coupled 
		rather strongly to its environment. Note the spin dimer in the blue circle.}
    \label{fig:strongISO}
\end{figure}

\end{widetext}

\end{document}